\documentstyle[11pt,newpasp,twoside,epsf,graphics,natbib]{article}

\newcommand{\nat}{Nature}

\begin{document}

\title{
Stellar Coronal Spectroscopy with the Chandra HETGS
} 

\author{
  David P.\ Huenemoerder, Bram Boroson, Norbert S. Schulz, Claude R. Canizares
}

\affil{
  Massachusetts Institute of Technology, 77 Vassar St., Cambridge, MA 02139
}
\author{
Derek L. Buzasi, Heather L. Preston
}
\affil{
U.S. Air Force Academy, Colorado Springs, CO 80840-6254
}

\author{
Joel H. Kastner
}
\affil{
Rochester Institute of Technology, Rochester, NY 14623
}

\begin{abstract}
  
  Spectroscopy with the Chandra High Energy Transmission Grating
  Spectrometer (HETGS) provides details on X-ray emission and activity
  from young and cool stars through resolution of emission lines from
  a variety of ions. We are beginning to see trends in activity
  regarding abundances, emission measures, and variability.  Here we
  contrast spectra of TV~Crt, a weak-lined T~Tauri star (WTT), with
  TW~Hya, a Classical T~Tauri star (CTT).  TV~Crt has a spectrum more
  like magnetic activity driven coronae, relative to the TW~Hya
  spectrum, which we have interpreted as due to accretion-produced
  X-rays.  We have also observed the long period system, IM~Pegasi to
  search for rotational modulation, and to compare activity in a long
  period active binary to shorter period systems and to the pre-main
  sequence stars. We detected no rotational modulation, but did see
  long-duration flares.
  
\end{abstract}

\section{Introduction}

X-ray emission is ubiquitous among late-type and pre-main sequence
(PMS) stars, as has been amply demonstrated by imaging and
low-resolution X-ray observatories \citep{Feigelson:Montmerle:1999}.
With the advent of the Chandra transmission-grating and the XMM-Newton
reflection-grating spectrometers, we can now probe the nature of the
X-ray emission in detail through high-resolution line and continuum
diagnostics. Early Chandra results confirmed some of the abundance
anomalies derived from low-resolution imaging spectra and also
unambiguously confirmed that few-component temperature models are
generally not adequate.  Much effort is now being spent to survey and
analyze stellar X-ray spectra for a range of intrinsic emission
conditions provided by evolutionary state, rotational periods, and
activity level.  We can derive temperature distributions and elemental
abundances from line-based emission measure analysis under assumptions
regarding collisional ionization equilibrium, ionization balance, and
uniformity of emitting plasma. Solutions are not unique even under
these conditions, but they are a necessary starting point.

Here we present preliminary results for TV~Crt (HD~98800), a WTT, and
IM~Pegasi (HD~216489, HR~8703), a long period RS~CVn binary.

\section{Pre-Main Sequence X-Ray Emission}

The TW~Hya Association (TWA) is an association of about 30
\citep{Zuckerman:Webb:2001} PMS stars at about 50 pc from the Sun
\citep{Kastner:Zuckerman:1997} and far from star forming clouds.  The
TWA has provided new insight into the process of star and planet
formation \citep[see for examples][]{Jayawardhana:Fisher:1998,
  Jura:Malkan:1998, Holland:Greaves:1998}.

We observed the WTT star system, TV~Crt (HD 98800), a TWA member, to
compare its X-ray spectrum with that of the actively accreting TW~Hya.
TV~Crt is a hierarchical quadruple and is one of the best studied
examples of a solar mass star-disk system \citep{Soderblom:King:1998}.
Two resolved sources, TV~Crt-$A$ and -$B$, are visible in the Chandra
zero-order image, with diskless component $A$ being about three times
brighter than $B$ \citep{Kastner:Huenemoerder:2003}.  Component $A$
also flared during our observation.  In the dispersed spectrum, we
cannot separate the $A\&B$ due to their proximity ($<1$ arcsec), given
the spectrograph's astigmatism in the spatial dimension.

The first high-resolution X-ray spectrum of a low-mass CTT, TW~Hya,
showed dramatic differences from other stellar coronal spectra
obtained to date: a very cool, sharply peaked temperature
distribution, and line ratios indicative of high densities.
Abundances were similar to coronae, but more extreme: very weak iron,
and very strong neon. Optical and IR observations of TW~Hya showed that
its disk is nearly face-on
\citep{Krist:Stapelfeldt:2000,Muzerolle:Calvet:2000,Trilling:Koerner:2001}.
From this and the peculiar X-ray line properties, we attributed the
X-ray emission to accretion rather than coronal emission from some
form of rotational dynamo-driven magnetic activity. \citet{Kastner:02}
described the results in detail.

\begin{figure}[htb]
\centering\leavevmode\scalebox{0.5}
  {\includegraphics{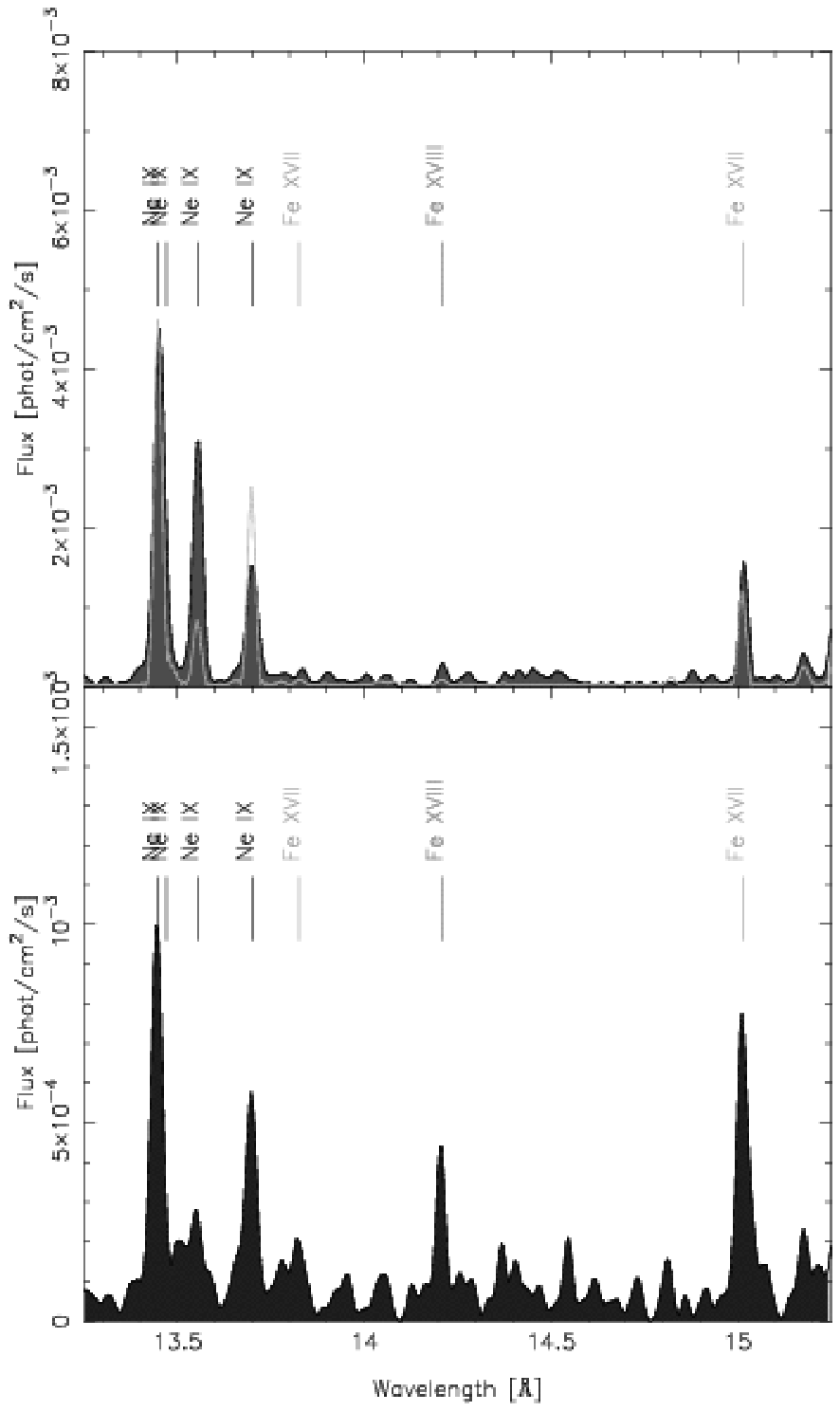}}
\centering\leavevmode\scalebox{0.5}
  {\includegraphics{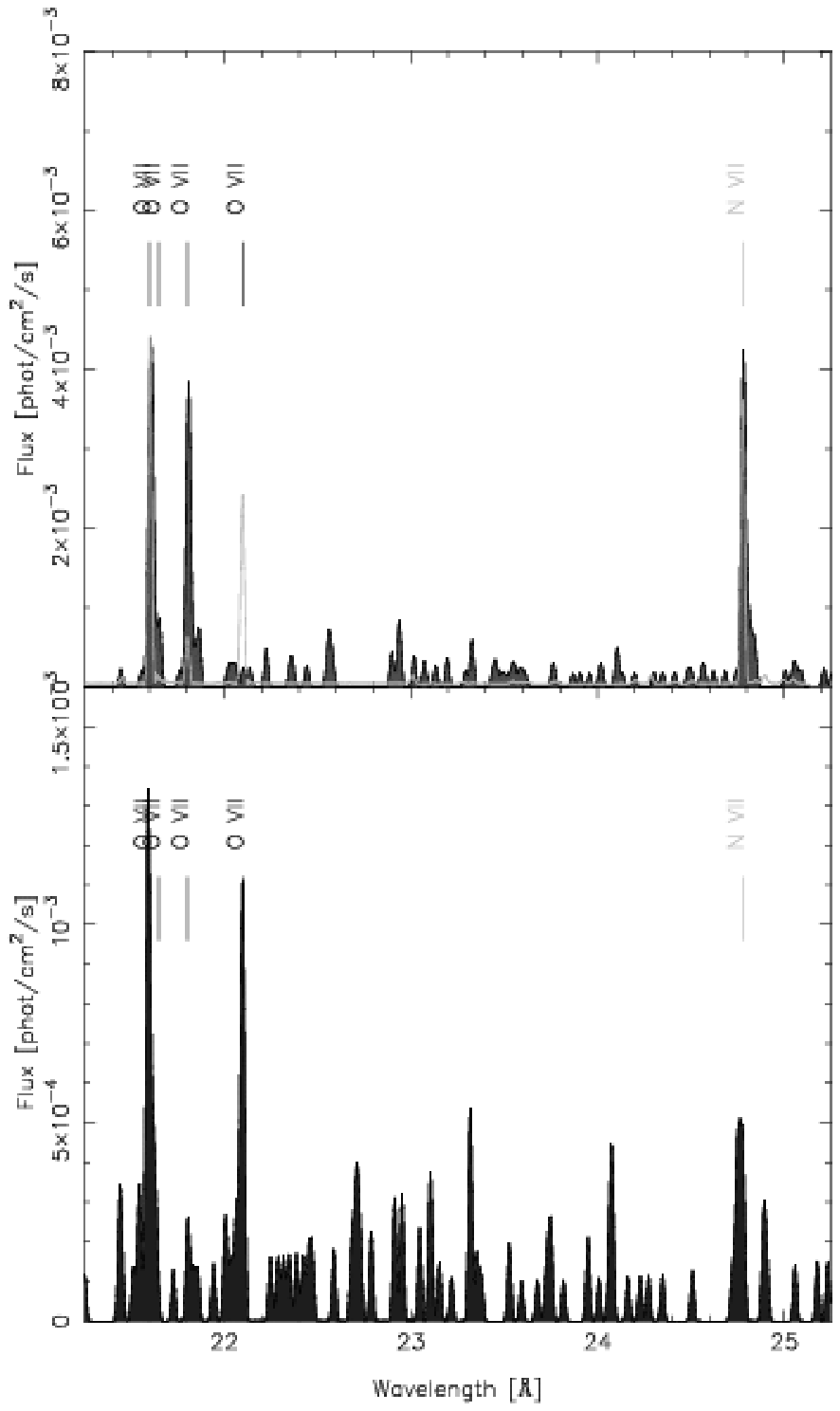}}
\caption{On the left is the Ne~{\sc ix}-triplet region, with resonance,
  intercombination, and forbidden lines at 13.447, 13.552, and 13.699
  \AA\ ($r$, $i$, $f$). The top graph is TW~Hya, and the bottom,
  TV~Crt.  Note the relative strengths of the $f$ and $i$ lines.  On
  the right is the O~{\sc vii} triplet region (21.6, 21.8, and 22.1
  \AA).  The difference is extreme: no detectable $f$ line in TW~Hya
  whereas in TV~Crt the $r$ and $f$ are of comparable intensity and
  the $i$ line is only a marginal detection (the gray, unfilled curve
  is the low-density theoretical spectrum).}
\label{fig:neotriplets}
\end{figure}

The difference between TV~Crt and TW~Hya in the helium-like,
density-sensitive, triplets of O~{\sc vii} and Ne~{\sc ix} is
obvious even in a casual inspection of the spectra.  Above
a critical density, collisional excitation dominates recombination of
the forbidden line ($f$) and increases the strength of the
intercombination line ($i$).  (The ratio can also be affected by
ultraviolet photo-excitation which is significant in systems with O or
B stars, but which we do not believe to be significant in TW~Hya.)
The $f/i$ ratios for TV~Crt are at or near the low density limit, more
like the typical coronal spectra of active binaries (presumably
magnetic-dynamo driven).  Figure~\ref{fig:neotriplets} shows the
Ne~{\sc ix} and O~{\sc vii} regions for each system.

Our preliminary emission measure modeling shows that TV~Crt's plasma
is hotter than that of TW~Hya (Figure~\ref{fig:dem}). Both systems
display a cool component at $\log T = 6.5$, but the emission measure
distribution of TV~Crt is flatter and remains high until about 7.0
before declining.  The abundance anomalies deduced from the TV~Crt
spectrum are not as extreme as in the case of TW~Hya, in which Fe and
Ne ratios to Solar were about 0.1 and 2.0; in TV~Crt they are about
0.2 and 1.0, respectively.  Further analysis requires iteration using
the derived plasma model to define a physical continuum, then
re-measuring lines, and re-deriving the emission measure and
abundances until consistent.

\section{Magnetically Active Binaries}

Solar activity has long been used as a paradigm for the interpretation
of activity in cool stars which are presumed to possess a magnetic
dynamo based on determinations of magnetic fields and activity cycles
\citep{Saar:1996,Baliunas:Nesme:1996}.  The existence of solar-like
features such as spots and flares in addition to correlations between
rotation rate and X-ray luminosity \citep{Pallavicini:1989} lends
strong support to the notion of stellar dynamos, but details of solar
and stellar dynamos remain obscure \citep[see, e.g.,][for a
discussion of dynamo theory]{Charbonneau:MacGregor:1997}.

One goal of research on stellar activity is to determine the
underlying coronal heating mechanism, which is not directly
observable.  Fundamental observable quantities must be derived for a
range of stellar parameters in order to provide the basis for a
theoretical understanding.  Our immediate goals are to use the phased
and total X-ray spectra of a long-period active binary to address several
issues: determine the temperature distribution; determine abundances;
probe the geometry via rotational modulation and density diagnostics;
monitor for flares; and compare to other stars vs.\ evolutionary state
and environment.

\begin{figure}[htb]
\centering\leavevmode\scalebox{0.7}
  {\includegraphics{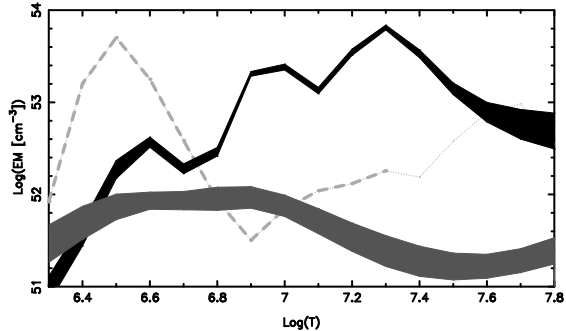}}
\caption{The upper filled curve is a differential emission measure fit to
  IM~Peg, and the lower to TV~Crt.  The breadth of the band is the one
  sigma dispersion from a Monte Carlo calculation in which each fit
  was obtained by perturbing the measured fluxes according to their
  statistical uncertainties. The dotted line is for TW~Hya, adapted from
  \citet{Kastner:02}, which is an upper limit above $\log T\sim7.3$.}
\label{fig:dem}
\end{figure}

IM~Pegasi, a 24d period, K2~{\sc iii}
RS~CVn binary, is one of a very few RS~CVn binaries to have shown
rotational modulation in both lower and upper atmospheric emission
\citep{Huenemoerder:Ramsey:1990}.  We monitored IM~Peg with the HETGS
during two rotations with eight observations of 25ks each.  While
there were slow flux modulations, they did not repeat with rotational
phase.  Hence, we presume they are signatures of long-duration flares,
especially since the amplitude of variability, though small, was
largest at short wavelengths.  We are in the process of performing
line-based, iterative emission measure analysis vs.\ flux level.  A
preliminary emission measure, integrated over the full 200ks is shown
in Figure~\ref{fig:dem}.  The emission measure and abundance
distributions for IM~Peg are quite similar to those of the 2-day
period RS~CVn, AR~Lac \citep{Huenemoerder:Canizares:al:2003}.
The hotter peak in the IM~Peg distribution (near $\log T=7.3$) is
probably due to flaring, since we have seen direct correlation of this
peak with flares in other stars \citep[e.g.,][]{Huenemoerder:01}.
Note the lack of such a peak in the TV~Crt curve, and the cool,
sharply peaked distribution of TW~Hya.

\section{Conclusions}

The comparison of three spectra from an actively accreting system, a
young but disk-free system, and a ``traditional'' magnetically active
corona shows a suggestive progression from accretion to dynamo driven
X-rays.  The temperature and density may be clues to differences in
X-ray origin.  But there are also similarities requiring explanation:
the elemental abundance trends, and the occurrence of flares.  A good
sample of high-resolution X-ray spectra of RS CVn binaries and flare
stars exists, but many more PMS spectra are needed to help resolve
these questions.

\acknowledgments

Support for this work was provided by NASA contract NAS8-01129, SAO
SV1-61010 (CXC/MIT), and NASA-Chandra Award Number G02-3005A issued by
the CXC, which is operated by SAO for and on behalf of NASA under
contract NAS8-39073. We also thank the Air Force Office of Scientific
Research for their support.



\end{document}